\pdfoutput=1

\documentclass[11pt]{article}

\usepackage[]{acl}

\usepackage{times}
\usepackage{latexsym}

\usepackage[T1]{fontenc}

\usepackage[utf8]{inputenc}

\usepackage{microtype}

\usepackage{inconsolata}

\usepackage{graphicx}

\usepackage{bm}
\usepackage{amsmath}
\usepackage{stfloats}
\usepackage{booktabs}
\usepackage{tabularx}
\usepackage{xcolor} 
\usepackage{bbding}

\title{GenSim: A General Social Simulation Platform with Large Language Model based Agents}

\author{Jiakai Tang$^{1}$, Heyang Gao$^1$, Xuchen Pan$^{3}$, Lei Wang$^{1}$, Haoran Tan$^1$, Dawei Gao$^3$, \\\textbf{Yushuo Chen}$^3$, \textbf{Xu Chen}$^1$\thanks{Corresponding author.}, \textbf{Yankai Lin}$^{1}$, \textbf{Yaliang Li}$^{3}$, \textbf{Bolin Ding}$^{3}$, \\ \textbf{Jingren Zhou}$^{3}$, \textbf{Jun Wang}$^{2}$, \textbf{Ji-Rong Wen}$^{1}$ \\
  $^1$Gaoling School of Artificial Intelligence, Renmin University of China\\ 
  $^2$ University College London \\
  $^3$Alibaba Group\\
  \texttt{}
}

\begin{document}
\maketitle
\begin{abstract}
With the rapid advancement of large language models (LLMs), recent years have witnessed many promising studies on leveraging LLM-based agents to simulate human social behavior. While prior work has demonstrated significant potential across various domains, much of it has focused on specific scenarios involving a limited number of agents and has lacked the ability to adapt when errors occur during simulation. To overcome these limitations, we propose a novel LLM-agent-based simulation platform called \textit{GenSim}, which:
(1) \textbf{Abstracts a set of general functions} to simplify the simulation of customized social scenarios;
(2) \textbf{Supports one hundred thousand agents} to better simulate large-scale populations in real-world contexts;
(3) \textbf{Incorporates error-correction mechanisms} to ensure more reliable and long-term simulations.
To evaluate our platform, we assess both the efficiency of large-scale agent simulations and the effectiveness of the error-correction mechanisms. 
To our knowledge, GenSim represents an initial step toward a general, large-scale, and correctable social simulation platform based on LLM agents, promising to further advance the field of social science. The relevant code and project are open-sourced on \url{https://github.com/TangJiakai/GenSim}.
\end{abstract}

\section{Introduction}
Social science, which focuses on human behavior, communication, and organization, is playing an increasingly significant role as world civilization advances. 
One important research paradigm in social science is collecting real human data. 
For instance, to study the effectiveness of positive psychology interventions,~\cite{bolier2013positive} recruited over 6,000 participants to observe their responses to controlled experiments. 
While the paradigm of collecting real human data is widespread in social science research, it suffers from significant drawbacks, such as high cost, poor controllability, and challenges in reproducibility, which have troubled researchers for a long time.

In the field of artificial intelligence (AI), researchers have discovered that language serves as a crucial carrier of intelligence~\cite{zhao2023survey}, and the objective of ``next-token prediction'' using a massive training corpus (\emph{i.e.}, large language models, LLMs) has the potential to achieve human-like intelligence.
With the advent of these high-intelligence models, a new ``AI for Social Science'' direction has emerged: leveraging LLMs as proxies for real humans to conduct social science experiments~\cite{gao2024large}. This approach provides the opportunities to fundamentally address the above challenges faced by social science research, potentially paving the way for an entirely new research paradigm.
For instance, Generative agents~\cite{park2023generative} leverages 25 agents to simulate human daily life, and finds that these agents can autonomously host parties and conduct mayoral election.
RecAgent~\cite{wang2023user} simulates user online behaviors, and studies the phenomenons of information cocoon and conformity behaviors.
EconAgent~\cite{li2024econagent} studies the macroeconomic behaviors using LLM-based agents in the context of dynamic markets.

\begin{figure*}[t]
\centering
\includegraphics[width=0.99\textwidth]{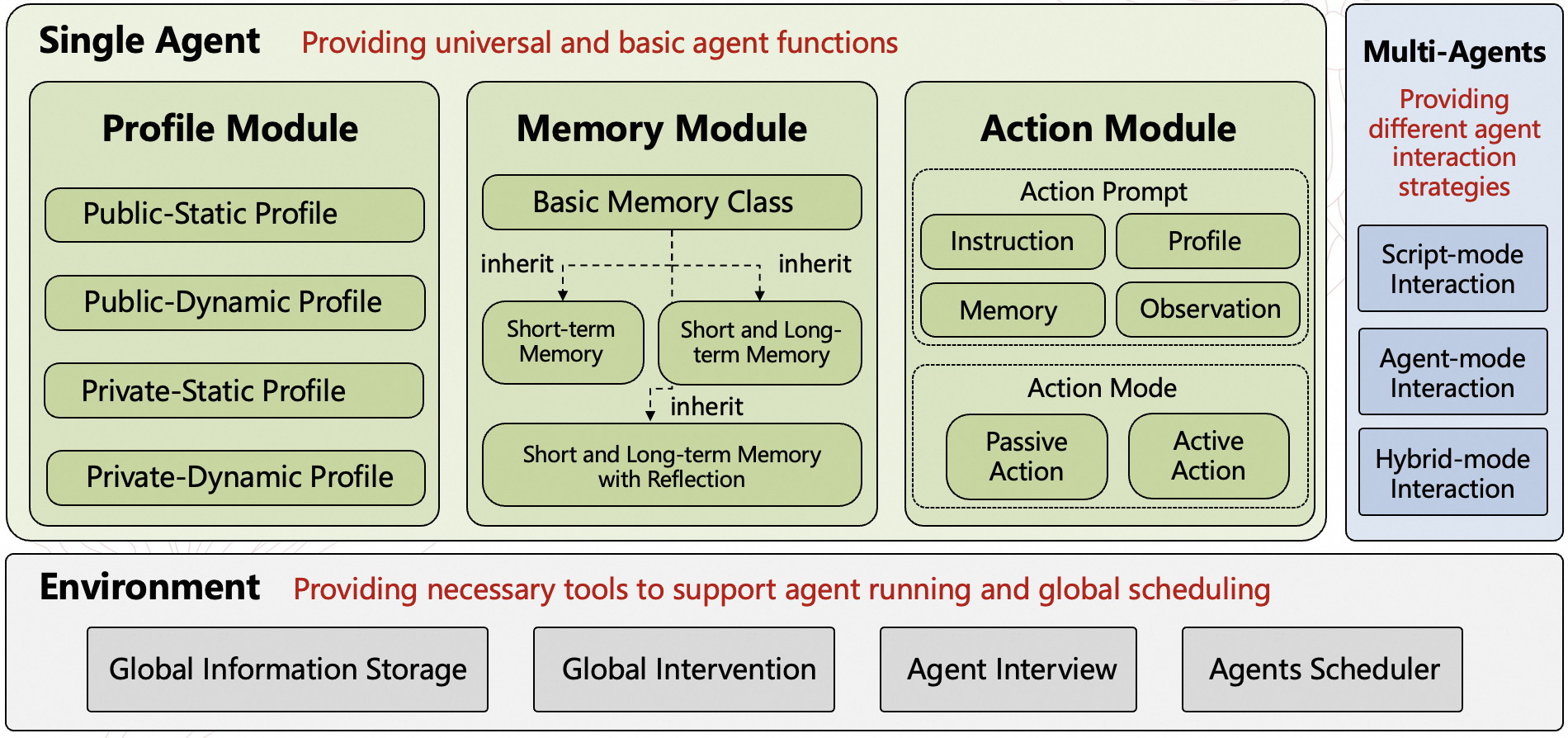}
\caption{The general framework for social simulation} 
\label{cxo}
\end{figure*}

While the above methods have shown promising results, they are primarily limited to specific scenarios and small-scale simulations. Moreover, when discrepancies arise between simulated behaviors and those observed in the real world, existing methods lack effective error-correction mechanisms. To address these limitations, we introduce GenSim, a general social simulation platform based on LLM agents.
In specific, to avoid reinventing the wheel in simulating various social scenarios, we propose a general programming framework composed of three key modules on single-agent construction, multi-agent scheduling, and environment setup. 
Additionally, we provide three default scenarios as references to help users quickly implement their customized simulations.
To achieve large-scale simulations in real-world scenarios, we leverage distributed parallel technology to support one hundred thousands of agents in our platform.
Finally, we design several error-correction mechanisms, allowing the platform to first perform self-evaluation or seek human feedback, and then fine-tune itself to ensure more reliable simulation. The comparison between our framework and the previous work can be seen in Figure~\ref{table:comparison}.

In summary, the main contributions of this paper are as follows:
(1) We propose a general, large-scale and correctable social simulation platform based on LLM agents.
(2) We provide detailed usage examples to illustrate the capabilities of our platform.
(3) We conduct a series of experiments to evaluate the platform's effectiveness and efficiency.

\section{Features of GenSim}
There are several unique features of our platform.
To begin with, we abstract a set of general functions to facilitate any customized simulation scenario according to the users' requirements.
Then, our platform supports one hundred thousand agents to better simulate large-scale populations in real-world contexts.
At last, we provide a series of error-correction mechanisms to ensure more reliable simulation.
The first two can be seen as static features from the generality and scalability perspectives, respectively, while the last one extends previous work from the dynamic perspective, making sure our platform can continually correct and improve itself.
\subsection{General Simulation Framework}
Our framework consists of three modules focusing on single agent, multi-agents, and environments (see Figure~\ref{cxo}). 
In the \textbf{single agent} module, users can flexibly configure the agent's profile, memory, and action components. The profile includes both public information, such as gender, name, and birthplace, as well as private attributes such as income and health condition. To enable the agent to retain behaviors in various ways, users can assemble different memory components like short-term memory, long-term memory, and the reflection mechanism to build the agent's memory.
The actions of the agents are driven by LLM prompts, where users can flexibly configure them to include agent profiles, memories and so on.

In the \textbf{multi-agents} module, inspired by the work~\cite{zhou2024real}, we design two strategies for generating agent interactions: script mode and agent mode. In specific, in script mode, all interactions are treated as a whole and generated in a single call to the LLM. For example, one can directly prompt LLMs to generate a dialogue between a doctor and a teacher in one step. In this strategy, the LLM acts as a meta-agent, producing the dialogue from a third-person perspective. In agent mode, interactions are generated by different agents, each representing a distinct role, and each agent generates outputs from a first-person perspective. This interaction generation process requires multiple calls to the LLM. In the example above, two agents are deployed, and each agent's output is determined by the complete history of their interactions.

\begin{figure*}[t]
\centering
\includegraphics[width=0.9\textwidth]{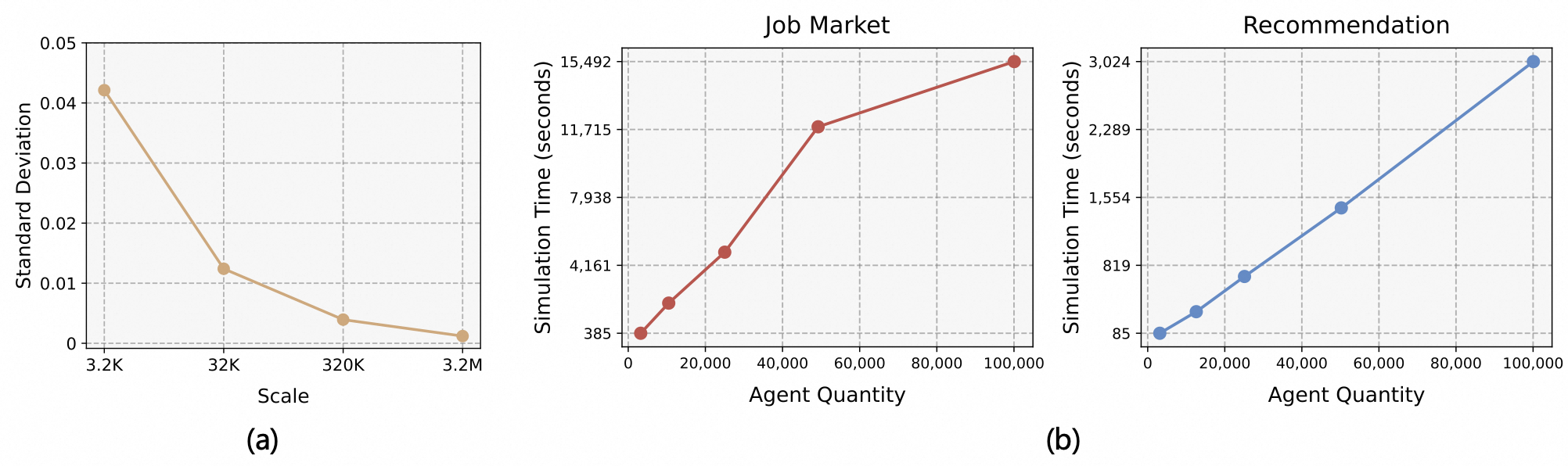}
\caption{(a) The fluctuation of the simulation results with different agent scales. 
(b) Time costs with different numbers of agents} 
\label{exp1}
\vspace{-0.4cm}
\end{figure*}

In the \textbf{environment} module, we store all the information beyond the agents necessary for running the simulation, such as the recommendation algorithm used in a web user simulator~\cite{wang2023user}. 
Additionally, we allow users to globally intervene in the platform, which is useful for counterfactual inferences. We also provide essential functions to facilitate interviewing, searching, and storing different agents.

Based on the above general framework, users can easily create customized simulations. 
To provide additional references, we offer three default scenarios: job market, recommender system, and group discussion. 
These scenarios can not only facilitate related research but also can provide code bases, enabling users to construct new scenarios with minimal effort.

\begin{figure*}[t]
\centering
\includegraphics[width=0.95\textwidth]{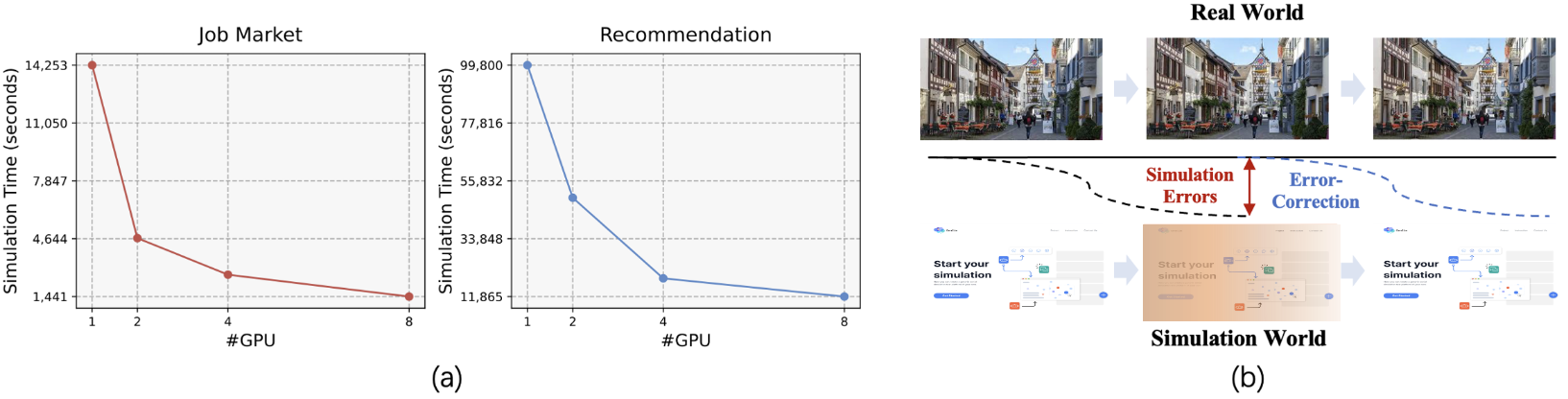}
\caption{(a) Time costs with different numbers of GPUs. 
(b) The error-correction mechanism.} 
\label{exp3}
\vspace{-0.2cm}
\end{figure*}

\subsection{Large-scale Simulation}
While there are many previous studies on leveraging LLMs to simulate human social behaviors~\cite{gao2024large}, the number of agents in their simulators are usually very small.
In such cases, the users need to sample a small set of individuals from the real-world large-scale populations, and then leverage agents to simulate the sampled individuals, assuming that these samples can accurately approximate real-world populations.
However, the sampled small number of individuals may lead to very large fluctuations of the simulation results.
To verify this statement, we conduct a preliminary experiment by simulating the user-item rating behaviors on a movie website.
In specific, we base our experiment on the well-known dataset MovieLens-32M~\cite{harper2015movielens}, which consists of 200,948 users' 32M ratings on 87,585 movies.
For each user-movie pair, we use LLMs to simulate the user's rating on the movie in the range of $\mathbf{R}=[0.5, 1.0, ..., 5.0]$.
To study the fluctuation of the simulation results with different agent scales, we first sample 3.2K, 32K, 320K and 3.2M user-item pairs from the complete dataset, and then, for each case, we repeat the simulation of predicting user-item ratings for 10 times.
Formally, suppose $\bm{p}_i$ represents the rating distribution of the $i$th experiment, where $i=1,2,…,10$. For each rating $r \in \mathbf{R}$, we compute the standard deviation across all experiments as
\begin{equation*}
    v(r) = \sigma(\bm{p}_1(r),\bm{p}_2(r),\dots,\bm{p}_{10}(r))
\end{equation*}
where $\sigma(\cdot)$ denotes the standard deviation operation. We use the sum of the standard deviations for all possible ratings to measure the fluctuation of the simulation results.
The experiment results are presented in Figure~\ref{exp1}(a), where we can see: as the number of samples becomes larger, the fluctuation of the simulation results is greatly lowered.
This result suggests that if we only have a small number of agents, then the simulation results can be not reliable, since it can be hardly reproduced due to the large simulation fluctuation.

\begin{figure}[t]
\centering
\includegraphics[width=\linewidth]{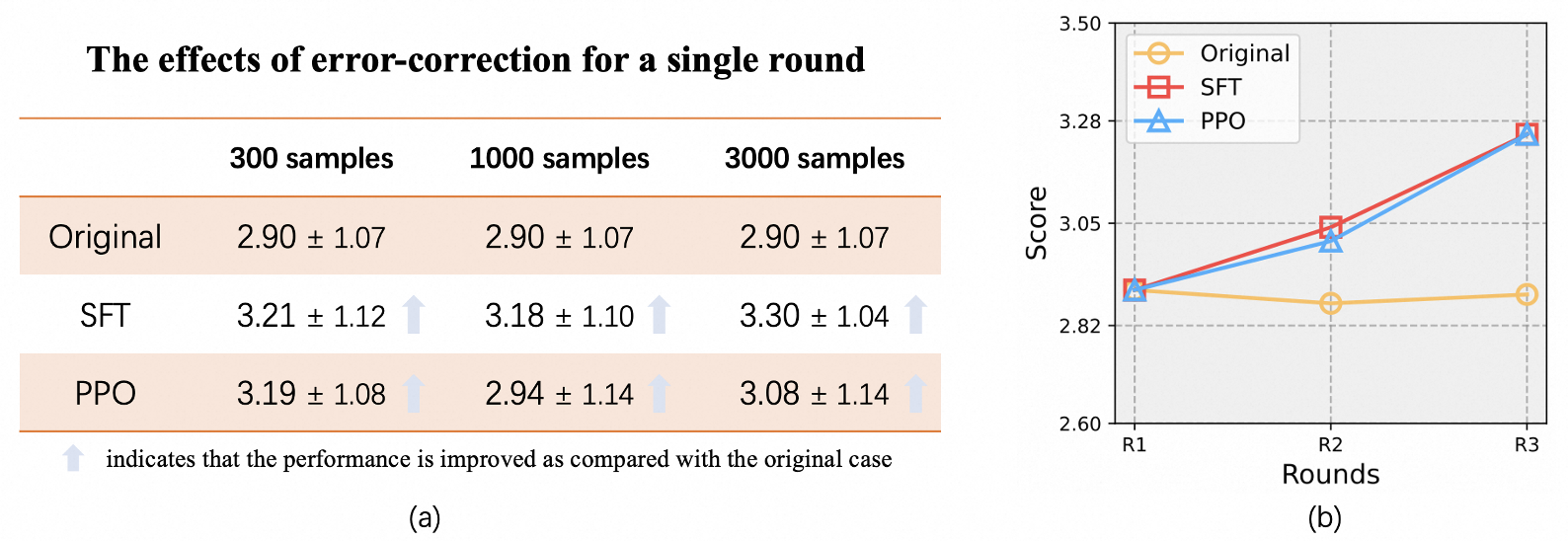}
\caption{The effects of our error-correction mechanisms in the job market scenario for both single round (a) and multi rounds (b) settings.} 
\label{exp2}
\end{figure}

To solve the above problem, in our platform, we support up to one hundred thousand agents to better simulate real-world scenarios.
To accelerate the simulation speed, we firstly employ an actor-based model~\cite{10.5555/1624775.1624804} to facilitate automatic parallel optimization. Specifically, the ``actor'' function acts as an independent unit that processes computations once it receives all the necessary messages. This method ensures that each agent, representing a participant, only performs computations when the essential input messages are available, thus enabling efficient parallel optimization. Furthermore, a dynamic workflow is designed to accommodate the probabilistic outputs of LLMs, which are not pre-determinable. This dynamic approach enhances flexibility, allowing the execution path to adapt based on variations in model inference. Last but not least, our distributed framework supports multi-machine parallel simulation, overcoming the limitations of single-machine simulation scalability.

Using proposed platform, we evaluate the simulation speed in the job market and recommendation scenarios, where we run our simulator for one round for both settings\footnote{For all experiments in this paper, we used a server with a 192-core CPU, eight A100-40G GPUs, and 440 GB of memory.}. All experiments in the paper use LLaMA3-8B as the foundational model for the agents.
The results are presented in Figure~\ref{exp1}(b), from which we can see: as the number of agents becomes larger, the time cost increases, and when we have 10w agents, they cost 15492 and 3024 seconds for running one round in the job market and recommendation scenarios, respectively.
In addition, we also evaluate the acceleration effects of distributed parallel computing in our platform.
In specific, we measure the time costs of running our platform for one round with different numbers of GPUs.
The results are presented in Figure~\ref{exp3}(a), where we can see, as the number of GPUs becomes larger, the time cost decreases, which suggests that, with the help of distributed parallel computing, our platform can effectively take the advantages of more GPUs.

\subsection{Simulation Error Correction}
\label{simulation_correction}
Most previous LLM-agent-based simulation platforms lack error-correction mechanisms, which means that if unexpected results occur during the simulation process, they can be accumulated and amplified as the simulation progresses (see Figure~\ref{exp3}(b)).
To solve this problem, in our platform, we provide two strategies for correcting the simulation errors.
The first one is based on LLMs, where we leverage GPT-4o to score on or revise the simulated result, utilizing the capabilities of LLMs as judges~\cite{zheng2023judging}.
The second one is based on real humans, where we provide interfaces for the users to score or revise the simulated agent behaviors.
Between these two approaches, the first is more efficient and requires no human intervention, though it may be less accurate due to the inherent biases of LLM. 
The second approach is more aligned with real humans but can be labor-intensive and less efficient.
Suppose the simulation result is represented by a $(q,a)$ pair, where $q$ is the prompt for driving an agent action, and $a$ is the action.
For each of the above strategies, there are two forms of feedback provided by LLMs or real humans.
Let the score for $(q,a)$ be $s$, and $a'$ be the revised results\footnote{It should be noted that $s$ and $a'$ may not co-exist for the same $(q,a)$ pair.}.
Then, we use $(q,a,s)$ and $(q, a')$ to fine-tune the backbone LLMs using PPO~\cite{schulman2017proximal} and SFT, respectively.

To evaluate the effectiveness of our designed error-correction mechanisms, we conduct experiments based on the job market scenario with LLMs as the feedback provider. 
To begin with, we evaluate whether PPO and SFT can improve the simulation results in a single round.
In the experiments, we select different numbers of samples for labeling, and use GPT-4o to measure the reasonableness of the simulation results, assigning scores from 1 to 5 based on reasonableness level, from low to high.
From Figure~\ref{exp2}(a), we can see: 
for both PPO and SFT, they can improve the simulation performance across different numbers of labeled samples.
Compared with PPO, the results of SFT are better, which is reasonable, since the revised action $a'$ used in SFT may include more effective and comprehensive information.

Next, we evaluate the effectiveness of the error-correction mechanisms in a multi-round setting. Specifically, we fine-tune the backbone LLMs in the earlier round and use the updated models to simulate the results in the subsequent round.
We present the results in Figure~\ref{exp2}(b).
We can see: if we do not conduct error-correction (the yellow line), the simulation performance is unsatisfactory.
When using PPO (the blue line) or SFT (the red line), the simulation performance improves significantly, and these improvements continue to increase as the number of simulation rounds grows.

\section{Usages of GenSim}
\begin{figure*}[t]
\centering
\includegraphics[width=\textwidth]{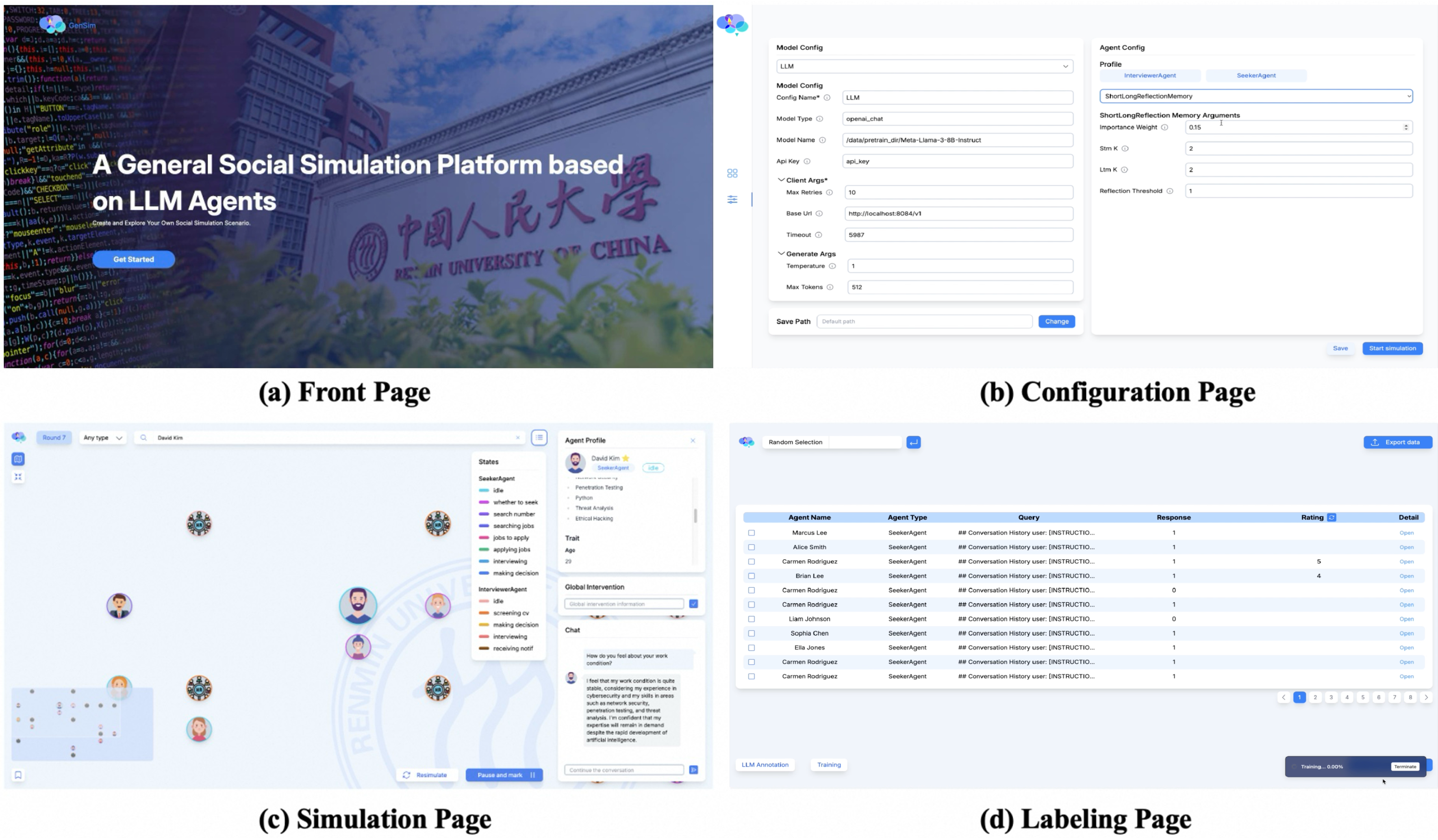}
\caption{The interfaces of GenSim.} 
\label{inter}
\end{figure*}

In this section, we introduce the basic methods for running a default scenario.
For more details, readers can refer to a short live demonstration of our platform on YouTube~\url{https://www.youtube.com/watch?v=SZf8mvhkLvI}.

The complete interfaces of our platform are presented in Figure~\ref{inter}. 
In specific, the user needs to click the `Get Started' button to initiate our platform.
The user can then configure the simulation by specifying scenarios, agent profile, memory type, number of agents, LLM parameters, etc. 
Once configured, the platform can be launched.
The simulation interface consists of three sections:
At the top, there is a search box that allows the user to search for an agent to observe its status and track its behavior.
In the middle, there is a display window where the behaviors of multiple agents are shown.
On the right, there is a functionality window where users can view agent profiles, intervene in the system, and interact with agents.
After running the platform for several rounds, the user can stop the simulation and label the results. Finally, the backbone LLMs can be fine-tuned based on the labeled results and used in the subsequent simulation.

\section{Related Work}
\subsection{Social Science}
Social science refers to the discipline that uses scientific methods to study a variety of social phenomena. It can be divided into different sub-disciplines such as political science, sociology, economics, psychology, and more. The social phenomena studied in social science are usually very sophisticated due to various uncertain variables, such as dynamically changing natural and social factors.

In the early stages, social science researchers conducted experiments in laboratory settings to conveniently control variables. For instance, \cite{Gosnell1926} studied the effect of non-partisan mail campaigns on increasing voter turnout by interviewing 6,000 individuals.
To enhance the reality and accuracy, \textit{Field Experiment} and \textit{Randomized Controlled Trial (RCT)} have gradually become the primary methods in social science research.
For example, ~\cite{staiano2014money} conducted a field experiment involving 60 participants over a period of six weeks, collecting various personal identity information. This helped in studying the monetary valuation of mobile personally identifiable information.
\cite{bolier2013positive} designed 39 randomized controlled studies involving 6,139 participants, aiming to evaluate the impact of positive psychology interventions on both the general public and individuals facing specific psychosocial issues.

Although existing social science experimental methods have been widely adopted, there are several significant drawbacks such as irreversibility, high cost, poor controllability, and even potential ethical violations. These issues severely restrict the feasibility and scalability of  social experiments. Moreover, long experimental durations and complex environmental factors may adversely impact the validity and timeliness of the experimental conclusions.
\subsection{Multi-Agent Simulation}
\begin{table*}[htbp]
\centering
\caption{An overview of comparing GenSim with existing works on multi-agent social simulations. The symbol `$\bm{\infty}$' indicates that our proposed platform can support general social simulation scenarios.}
\begin{tabularx}{0.88\textwidth}{>{\centering\arraybackslash}m{3cm}>{\centering\arraybackslash}m{4.5cm}>{\centering\arraybackslash}m{2.5cm}>{\centering\arraybackslash}m{2.5cm}}
\toprule
Name             & Simulation Domain & Agent Number & Self-evolution \\
\midrule
Generative Agent &        Daily Life           &        25      &   \textcolor{red}{\textbf{\XSolidBrush}}             \\
RecAgent         &      Recommendation Systems             &      20        &       \textcolor{red}{\textbf{\XSolidBrush}}         \\
EconAgent        &        Economic Market           &       100       &         \textcolor{red}{\textbf{\XSolidBrush}}       \\
Social Simulacra &         Social Network          &      1,000        &        \textcolor{red}{\textbf{\XSolidBrush}}        \\
Agent Hospital   &          Healthcare         &     <100         &   \textcolor{red}{\textbf{\XSolidBrush}}             \\
WarAgent         &         Warfare          &      <100        &    \textcolor{red}{\textbf{\XSolidBrush}}            \\
\hline
\addlinespace[0.4ex]
\textbf{GenSim (ours)}           &     $\bm{\infty}$             &     \textbf{>100,000}         & \textcolor{green}{\CheckmarkBold} \\          
\bottomrule
\end{tabularx}
\label{table:comparison}
\end{table*}
With the rise of large language models~\cite{achiam2023gpt,team2023gemini}, the "AI for Social Sciences" direction presents unprecedented opportunities for researchers in both social sciences and artificial intelligence. The core idea is to utilize LLMs as human-like brains to create behavior decision-makers that can approximately imitate real individuals, known as LLM-based agents.

Recently, there are increasing amount of works focused on how to leverage LLM-based multi-agents to conduct social simulation experiments in various fields. Specifically, Generative Agent~\cite{park2023generative}, as a pioneering work in this field, simulates the daily life of 25 agents in a small town. Similarly, RecAgent~\cite{wang2023user} and Social Simulacra~\cite{park2022social} investigate social phenomena that may emerge from group behavior in recommendation systems and social networks, such as information cocoons and the spread of antisocial behavior. In addition, EconAgent~\cite{li2024econagent}, Agent Hospital~\cite{li2024agent}, and WarAgent~\cite{hua2023war} each use multi-agent simulation in the fields of economic markets, healthcare, and warfare respectively, to model specific human behaviors in their specific domains, and further analyze individual actions and group social phenomena.

While the aforementioned methods demonstrate the powerful advantages of LLM-based agents in social simulation, there are still three significant shortcomings hindering the further development: existing works cannot support different domains and large-scale social simulations, and lack a error correction mechanism. In contrast, our framework effectively overcomes these issues. We can support general-domain simulations and large-scale multi-agents, and introduce a novel self-correcting self-evolution mechanism, achieving more reliable and accurate simulation performance. Overall comparison results are summarized in Table~\ref{table:comparison}.

\section{Conclusions}  
In this paper, we introduce a general, large-scale, and correctable social simulation platform based on LLM agents.
This is the initial version of our platform, we believe there is still much room left for improvement.
In the future, we plan to incorporate more advanced simulation accelerating strategies, and develop more adaptive self-correction mechanisms to improve the simulation performance.


\section*{Limitation}
Despite our pioneering effort in developing a general social simulation platform using LLM-based agents, there are still several limitations. \textbf{First}, the platform's generalization capability across diverse sociocultural contexts has not been fully verified. While agent simulations are more cost-effective than real-world experiments, their accuracy in modeling complex cultural norms, regional socioeconomic dynamics, and varied institutional settings requires further validation through comparative studies with traditional experimental approaches. \textbf{Second}, our reliance on LLM-as-a-judge for synthetic data evaluation introduces potential assessment biases. The prohibitive cost of human annotation prevented systematic verification of alignment between LLM-generated scores and expert human ratings, particularly in culturally sensitive scenarios where language models may encode latent value misalignments. This necessitates future research on calibration techniques for automated social evaluation metrics. \textbf{Third}, the self-correction mechanisms of LLMs in social simulations present unresolved reliability concerns. The inherent biases in LLMs' error identification and correction processes across varied social scenarios (especially those involving ethical dilemmas or conflicting group interests) remain unexamined. We leave these limitations as our future work to explore.

\section*{Ethical Considerations}
The data used in this paper comes from public datasets or simulated experiments using synthetic data generated by Large Language Models (LLMs). For public datasets, we adhere to their usage licenses to ensure that our work does not present any ethical issues.

\section*{Acknowledgments}
This work is supported in part by National Natural Science Foundation of China (No. 62422215 and No. 62472427), Beijing Outstanding Young Scientist Program NO.BJJWZYJH012019100020098, Intelligent Social Governance Platform, Major Innovation \& Planning Interdisciplinary Platform for the “DoubleFirst Class” Initiative, Renmin University of China, Public Computing Cloud, Renmin University of China, fund for building world-class universities (disciplines) of Renmin University of China, Intelligent Social Governance Platform.

\bibliography{main}

\begin{thebibliography}{18}
\providecommand{\natexlab}[1]{#1}

\bibitem[{Achiam et~al.(2023)Achiam, Adler, Agarwal, Ahmad, Akkaya, Aleman, Almeida, Altenschmidt, Altman, Anadkat et~al.}]{achiam2023gpt}
Josh Achiam, Steven Adler, Sandhini Agarwal, Lama Ahmad, Ilge Akkaya, Florencia~Leoni Aleman, Diogo Almeida, Janko Altenschmidt, Sam Altman, Shyamal Anadkat, et~al. 2023.
\newblock Gpt-4 technical report.
\newblock \emph{arXiv preprint arXiv:2303.08774}.

\bibitem[{Bolier et~al.(2013)Bolier, Haverman, Westerhof, Riper, Smit, and Bohlmeijer}]{bolier2013positive}
Linda Bolier, Merel Haverman, Gerben~J Westerhof, Heleen Riper, Filip Smit, and Ernst Bohlmeijer. 2013.
\newblock Positive psychology interventions: a meta-analysis of randomized controlled studies.
\newblock \emph{BMC public health}, 13:1--20.

\bibitem[{Gao et~al.(2024)Gao, Lan, Li, Yuan, Ding, Zhou, Xu, and Li}]{gao2024large}
Chen Gao, Xiaochong Lan, Nian Li, Yuan Yuan, Jingtao Ding, Zhilun Zhou, Fengli Xu, and Yong Li. 2024.
\newblock Large language models empowered agent-based modeling and simulation: A survey and perspectives.
\newblock \emph{Humanities and Social Sciences Communications}, 11(1):1--24.

\bibitem[{Gosnell(1926)}]{Gosnell1926}
Harold~F. Gosnell. 1926.
\newblock An experiment in the stimulation of voting.
\newblock \emph{American Political Science Review}, 20(4):869--874.

\bibitem[{Harper and Konstan(2015)}]{harper2015movielens}
F~Maxwell Harper and Joseph~A Konstan. 2015.
\newblock The movielens datasets: History and context.
\newblock \emph{Acm transactions on interactive intelligent systems (tiis)}, 5(4):1--19.

\bibitem[{Hewitt et~al.(1973)Hewitt, Bishop, and Steiger}]{10.5555/1624775.1624804}
Carl Hewitt, Peter Bishop, and Richard Steiger. 1973.
\newblock A universal modular actor formalism for artificial intelligence.
\newblock In \emph{Proceedings of the 3rd International Joint Conference on Artificial Intelligence}, IJCAI'73, page 235–245, San Francisco, CA, USA. Morgan Kaufmann Publishers Inc.

\bibitem[{Hua et~al.(2023)Hua, Fan, Li, Mei, Ji, Ge, Hemphill, and Zhang}]{hua2023war}
Wenyue Hua, Lizhou Fan, Lingyao Li, Kai Mei, Jianchao Ji, Yingqiang Ge, Libby Hemphill, and Yongfeng Zhang. 2023.
\newblock War and peace (waragent): Large language model-based multi-agent simulation of world wars.
\newblock \emph{arXiv preprint arXiv:2311.17227}.

\bibitem[{Li et~al.(2024{\natexlab{a}})Li, Wang, Zhang, Li, Lai, Kang, Ma, and Liu}]{li2024agent}
Junkai Li, Siyu Wang, Meng Zhang, Weitao Li, Yunghwei Lai, Xinhui Kang, Weizhi Ma, and Yang Liu. 2024{\natexlab{a}}.
\newblock Agent hospital: A simulacrum of hospital with evolvable medical agents.
\newblock \emph{arXiv preprint arXiv:2405.02957}.

\bibitem[{Li et~al.(2024{\natexlab{b}})Li, Gao, Li, Li, and Liao}]{li2024econagent}
Nian Li, Chen Gao, Mingyu Li, Yong Li, and Qingmin Liao. 2024{\natexlab{b}}.
\newblock Econagent: large language model-empowered agents for simulating macroeconomic activities.
\newblock In \emph{Proceedings of the 62nd Annual Meeting of the Association for Computational Linguistics (Volume 1: Long Papers)}, pages 15523--15536.

\bibitem[{Park et~al.(2023)Park, O'Brien, Cai, Morris, Liang, and Bernstein}]{park2023generative}
Joon~Sung Park, Joseph O'Brien, Carrie~Jun Cai, Meredith~Ringel Morris, Percy Liang, and Michael~S Bernstein. 2023.
\newblock Generative agents: Interactive simulacra of human behavior.
\newblock In \emph{Proceedings of the 36th annual acm symposium on user interface software and technology}, pages 1--22.

\bibitem[{Park et~al.(2022)Park, Popowski, Cai, Morris, Liang, and Bernstein}]{park2022social}
Joon~Sung Park, Lindsay Popowski, Carrie Cai, Meredith~Ringel Morris, Percy Liang, and Michael~S Bernstein. 2022.
\newblock Social simulacra: Creating populated prototypes for social computing systems.
\newblock In \emph{Proceedings of the 35th Annual ACM Symposium on User Interface Software and Technology}, pages 1--18.

\bibitem[{Schulman et~al.(2017)Schulman, Wolski, Dhariwal, Radford, and Klimov}]{schulman2017proximal}
John Schulman, Filip Wolski, Prafulla Dhariwal, Alec Radford, and Oleg Klimov. 2017.
\newblock Proximal policy optimization algorithms.
\newblock \emph{arXiv preprint arXiv:1707.06347}.

\bibitem[{Staiano et~al.(2014)Staiano, Oliver, Lepri, De~Oliveira, Caraviello, and Sebe}]{staiano2014money}
Jacopo Staiano, Nuria Oliver, Bruno Lepri, Rodrigo De~Oliveira, Michele Caraviello, and Nicu Sebe. 2014.
\newblock Money walks: a human-centric study on the economics of personal mobile data.
\newblock In \emph{Proceedings of the 2014 ACM International Joint Conference on Pervasive and Ubiquitous Computing}, pages 583--594.

\bibitem[{Team et~al.(2023)Team, Anil, Borgeaud, Alayrac, Yu, Soricut, Schalkwyk, Dai, Hauth, Millican et~al.}]{team2023gemini}
Gemini Team, Rohan Anil, Sebastian Borgeaud, Jean-Baptiste Alayrac, Jiahui Yu, Radu Soricut, Johan Schalkwyk, Andrew~M Dai, Anja Hauth, Katie Millican, et~al. 2023.
\newblock Gemini: a family of highly capable multimodal models.
\newblock \emph{arXiv preprint arXiv:2312.11805}.

\bibitem[{Wang et~al.(2023)Wang, Zhang, Yang, Chen, Tang, Zhang, Chen, Lin, Song, Zhao et~al.}]{wang2023user}
Lei Wang, Jingsen Zhang, Hao Yang, Zhiyuan Chen, Jiakai Tang, Zeyu Zhang, Xu~Chen, Yankai Lin, Ruihua Song, Wayne~Xin Zhao, et~al. 2023.
\newblock User behavior simulation with large language model based agents.
\newblock \emph{arXiv preprint arXiv:2306.02552}.

\bibitem[{Zhao et~al.(2023)Zhao, Zhou, Li, Tang, Wang, Hou, Min, Zhang, Zhang, Dong et~al.}]{zhao2023survey}
Wayne~Xin Zhao, Kun Zhou, Junyi Li, Tianyi Tang, Xiaolei Wang, Yupeng Hou, Yingqian Min, Beichen Zhang, Junjie Zhang, Zican Dong, et~al. 2023.
\newblock A survey of large language models.
\newblock \emph{arXiv preprint arXiv:2303.18223}.

\bibitem[{Zheng et~al.(2023)Zheng, Chiang, Sheng, Zhuang, Wu, Zhuang, Lin, Li, Li, Xing et~al.}]{zheng2023judging}
Lianmin Zheng, Wei-Lin Chiang, Ying Sheng, Siyuan Zhuang, Zhanghao Wu, Yonghao Zhuang, Zi~Lin, Zhuohan Li, Dacheng Li, Eric Xing, et~al. 2023.
\newblock Judging llm-as-a-judge with mt-bench and chatbot arena.
\newblock \emph{Advances in Neural Information Processing Systems}, 36:46595--46623.

\bibitem[{Zhou et~al.(2024)Zhou, Su, Eisape, Kim, and Sap}]{zhou2024real}
Xuhui Zhou, Zhe Su, Tiwalayo Eisape, Hyunwoo Kim, and Maarten Sap. 2024.
\newblock Is this the real life? is this just fantasy? the misleading success of simulating social interactions with llms.
\newblock \emph{arXiv preprint arXiv:2403.05020}.

\end{thebibliography}

\clearpage

\end{document}